\documentclass[aps, preprint, showpacs, epsfig]{revtex4}
\usepackage{graphics}
\usepackage{graphicx}
\usepackage{epsfig}
\def \be {\begin{equation}}
\def \ee {\end{equation}}
\def \ba {\begin{eqnarray}}
\def \ea {\end{eqnarray}}
\def \bm {\begin{displaymath}}
\def \em {\end{displaymath}}

\begin{document}
\title{ Partition function of a bubble formed in double stranded DNA}
\author{ Yashwant Singh}
\affiliation{Department of Physics, Banaras Hindu University,
Varanasi-221 005,
India}

\date{\today}
\begin{abstract}
We calculate the entropic part of partition function of a bubble 
embedded in a double stranded DNA (dsDNA) by 
considering the total weights of possible configurations of a system
of two single stranded DNA (ssDNA) of given length which start 
from a point along the contour of 
dsDNA and reunite at a position vector {\bf r} measured from the first
point and the distribution function of the position vector {\bf r} 
which separates the two zipper forks of the bubble 
in dsDNA. For the distribution function of position vector {\bf r}
we use the distribution of the end-to-end vector {\bf r} of 
strands of given length of dsDNA found from
the wormlike chain model. We show that when the chains forming the bubble 
are assumed to be Gaussian the so called loop closure
exponent $c$ is 3
and when we  made correction by including self avoidence in each chain
the value of $c$ becames 3.2.

\end{abstract}

\maketitle
DNA is a set of two entangled polymers connected by hydrogen bonds
between complementary bases and base-stacking between nearest
neighbour pairs of base pairs and wound around each other to make a helix \cite{1}.
Though a very long  DNA behaves like a flexible polymer in a solution,
at smaller length scales (which may be of the order of hundreds of base-pairs.)
it exhibits considerable stiffness. The stiffness of a polymer chain is measured by the
persistent
length $\l_p$, the length scale at which the tangent vectors to the polymer curves
are decorrelated \cite{2}. On scale smaller than $\l_p$, bending energy dominates and the
chain is relatively stiff with little or no bending. The $l_p$ for the double-stranded
DNA (dsDNA) is experimentally known to be about $500\AA$ which is equivalent
to $145$ bases long along its contour \cite{3}. In comparison, the persistent length
for single stranded DNA (ssDNA) depends sensitively on the ionic strength of the
solvent and can be as short as $10\AA$ \cite{4}.

In physiological solvent conditions the average value of the interactions
for a base-pair that stabilize the dsDNA structure is of the order of
few $k_BT$ (thermal energy) \cite{5} and thermal fluctuations can lead to local
and transitory unzipping of the double-strands \cite{6,7}. The cooperative
opening of a sequence of consecutive base pairs leads to formation of local
denaturation zones (bubbles). A DNA bubble consists of flexible single-stranded DNA
and its size fluctuate by zipping and unzipping of base-pairs at the two zipper
forks (points $P_1$ and $P_2$ in Fig. 1(b)) where the bubble connects 
the intact double-strands. As the persistent
length $l_s$ of the ssDNA is several times smaller than that of dsDNA the two chains fluctuate
relatively freely and make entropic contributions to free energy of the bubble. The existence
of bubbles in dsDNA represents the classic competition between energy and entropy; the intact
dsDNA regions are dominated by the interaction energy due to hydrogen bonding of pairs and base
stacking, bubbles by the entropy gain on disruption of base-pairs.
The average size of a bubble depends on the sequence of base-pairs, temperature
and ionic strength and varies from few broken base-pairs at room temperature
to few hundreds open base pairs close to melting temperature $T_m$ [8,9].
Below $T_m$, once formed a bubble is intermittant feature and eventually zip
close again. This DNA breathing can be probed on the single
molecule level in real time by fluorosence method \cite{7}. Above
$T_m$ individual bubbles continuously increase in size and
merge with vicinal bubbles until complete denaturation \cite{10}.

In the study of thermal denaturation of dsDNA [10,11,12] as well as in the
study of static and dynamic properties of an intermittant bubble [13,14] 
one needs to know its free energy. The partition function of a bubble
 of length $l_b$ has been approximated by the number of
configurations of walks of $2l_b$ length returning for the first time to the
origin \cite{15} which in the limit $l_b\rightarrow \infty$ assumes the
following form \cite{16}

\ba
Z(l_b)\sim \frac{\sigma{\mu}^{2l_b}}{(2l_b)^c}
\ea
where $\mu$ is a non-universal geometric factor while $\it{c}$, so called
loop closure exponent, is a universal quantity. We refer the model leading
to (1) as closed walks model. The factor $\sigma$ is used as an adjustable
parameter to fit the denaturation curves of DNA and its value has been found to
depend on the value of exponent $\it{c}$ \cite{12}. For example for $\it{c}$=1.75
the value of $\sigma$ that fits the experimental data is equal to $1.26\times 10^{-5}$
whereas for $\it{c}$=2.15, its value is found to be ten times larger 
$(i.e. \sigma =1.26\times 10^{-4})$
. Though $\sigma$ is called cooperativity parameter, its use is 
essentially  to correct the configurations of the bubble which have been
overestimated by the closed walks model.

For closed random walks embedded in three dimensional space (3-d) the exponent
$\it{c}$ has been found to be $1.5$. The inclusion of the excluded volume
interactions within the bubble gives $\it{c}$=1.76 \cite{15}. When the
excluded volume interactions between a bubble and the rest of the chain
is taken, then $\it{c}\simeq$ 2.11 \cite{17,18}. The nature of the melting
transition depends on the value of $\it{c}$ \cite{10}. For $\it{c}\leq1$
there is no transition, for $1< {\it{c}}\leq2$, the transition is continuous
while for $c>2$, the transition is first 
order.

The closed walks model which lead to Eq.(1) with values of $\it{c}$ given above
 for the partition function of a bubble, however, ignores the fact that the bubble is embedded
in a dsDNA which imposes constraints on the distribution function of vector $\bf r $
connecting the two ends (zipper forks) of the bubble. In this Letter we show that 
 when this constraint is taken into account the value of $\it{c}$ becomes greater
than 2 even when the bubble is assumed to be formed by two non-interacting Gaussian chains.

In Fig.1(a) we show a segment of dsDNA in which points $P_1$ and $P_2$ are separated by distance
$l_b=nb_0$  along the contour of the chain. Here $n$ is the number of base-pairs between $P_1$ and $P_2$
and $b_0=3.4\AA$, the average distance between successive base-pairs in the Watson-Crick
model of dsDNA \cite{1}. In space the points $P_1$ and $P_2$ are separated by vector $\bf r$.
In Fig.1(b) we show a bubble which is formed by unzipping of all n base-pairs between $P_1$
and $P_2$. The points $P_1$ and $P_2$ represent the two zipper-forks of the bubble. 
Each of the two
chains $C_1$ and $C_2$ of the bubble is of length $l_0=na_0$ where $a_0\simeq 6\AA$.
To calculate the entropic part of partition function of the bubble of 
length $l_b$ formed by unzipping of $n$ base-pairs of dsDNA we use the relation 
\ba
Z_n(l_b)=\int d{\bf r}Z_n({\bf r},l_0)\rho ({\bf r},l_b)
\ea

where $Z_n({\bf r},l_0)$ are the total weights of the possible 
configurations of a system of two ssDNA each of length
$l_0$ which start at origin and reunite through hydrogen
bonds and stacking interactions at a point defined 
by position vector ${\bf r}$  and $\rho({\bf r},l_b)$
is the normalized probability density of position
vector ${\bf r}$ separating the two zipper forks 
of the bubble in dsDNA. As we show below the
  factor $\rho({\bf r},l_b)$ which
has been ignored in the closed walks model plays
crucial role in determining the total number
of configurations of a bubble.

If we assume each chain forming the bubble to be Gaussian
, its distribution for the end-to-end  vector for $l_0 >>l_s$ can be written as \cite{2}
\ba
P_n({\bf r},l_0)=\left(\frac{4\pi}{3}l_0l_s\right)^{-3/2}exp
\left(\frac{-3r^2}{4l_0l_s}\right)
\ea
where $2l_s$ is taken to be equal to Kuhn statistical segment length. All distances
here and below are expressed in unit of $b_0$ and therefore made dimensionless.
The total weights of the possible configurations of two random walks starting
from point $P_1$ (origin) and reuniting at a distance $r$ at
point $P_2$ are therefore
\ba
Z_n({\bf r},l_0)=\left(\frac{4\pi}{3}l_0l_s\right)^{-3}
exp\left(\frac{-3r^2}{2l_0l_s}\right)
\ea

When one substitutes Eq.(4) in Eq.(2) and chooses $\rho({r,l_b})$=1 one gets 
$Z_n(l_b)\propto{\large{1/{l_b}^c}}$
with $\it {c}$ = 1.5 in agreement with the result 
reported in literature for closed random walks.
Note that choosing $\rho({r,l_b})$=1 amounts to assuming that the vector $\bf r $
separating the zipper forks of the bubble of length
$l_b$ in a dsDNA can have values between zero and
infinity with equal probability which is obviously incorrect as it ignores
the fact that the bubble is embedded in a dsDNA and its ends separation
is constrained by the dsDNA strands.
As shown in Fig.1(b) the zipper forks $P_1$  of the bubble on its left
and the fork $P_2$ on its right are connected to long strands of dsDNA. The
fluctuational motion of these forks in space (not along the dsDNA chain
 which changes the value of $l_b$) will therefore be controlled by these
strands. Here we are considering  dsDNA of contour length L$\rightarrow\infty$
, $l_b/L\rightarrow 0$ and the bubble located away from the two ends
of dsDNA . For a dsDNA which contour length is of the order of $l_p$ or less
the bubble formation may  change the conformational behaviour
of the chain  [26,27], but for L$\rightarrow\infty$ such change is expected
to be negligible as any change in their position amounts to moving or
rotating the dsDNA segments attached to these points on either or both
sides which may cost huge energy. It therefore seems reasonable
to assume that the distribution of distance $r$ separating points $P_1$
and $P_2$ in Fig.1(a) remains unchanged after the formation of the
bubble and $\rho({\bf r},l_b)$ can be approximated by 
 the probability density of finding points $P_1$ and $P_2$ at separation 
${\bf r}={\bf r}(s)-{\bf r}(s)^\prime$ where $l_b=s-s^\prime$ is the 
distance between $P_1$ and $P_2$ along the contour of dsDNA.

In order to understand the effect of possible approximation to $\rho({r,l_b})$ it may be
useful to consider some limiting cases : In the case of dsDNA being a rigid rod
of infinite length, $\rho({r,l_b})$ can be given as $\rho({r,l_b})=
\delta(r-l_b)/4\pi{l_b}^2$ as positions of $P_1$ and $P_2$ remain unchanged due to
formation of the bubble. For this case the partition function $Z_n(l_b)$ reduces to
\ba
Z_n(l_b)=\frac{K_1(\mu_G)^{l_b}}{l_b^3}
\ea
where

$\hspace{2cm}${\large$K_1=(\frac{3}{4\pi}\frac{1}{a_0l_s})^3$} 
$\hspace{1cm}$and  $\hspace{2cm}$
{\large$\mu_G=e^{-\frac{3}{2a_0l_s}}$}

In another limit of dsDNA being a freely joined phantom chain, $\rho({r,l_b})$
can be found from the end-to-end distribution function of a Gaussian chain
of length $l_b >>l_p$. Thus \cite{19}
\ba
\rho(r,l_b)=\left(\frac{4\pi l_bl_p}{3}\right)^{-3/2}exp\left(\frac
{-3r^2}{4l_bl_p}\right)\nonumber\\
\ea

where Kuhn statistical segment length is equal to $2l_p$. Substituting
this in Eq.(2) we get
\ba
Z_n(l_b)=\frac{K_2}{l_b^3} ,\hspace{2cm} K_2=\left(\frac{3}{4\pi}\right)^3
\left(\frac{1}{2a_0l_sl_p+{a_0}^2{l_s}^2}\right)^{3/2}
\ea

In both limits we find $\it{c}$=3.

A dsDNA being a semi-flexible polymer, its conformational properties
can be found from a wormlike chain model in which the polymer is
represented by a differential space curve ${\bf r}(s)$ of length
L parametrized to  arc length \cite{20}. The model is specified by
the Hamiltonian,
\ba
\beta H=\frac{1}{2}l_p{\int_0}^L ds\left(\frac{d{\bf\hat {t}}(s)}{ds}\right)^2
\ea

where {\large$ { \bf\hat{t}}(s)=\frac{d{\bf r}(s)}{ds}$} is the unit tangent
vector to the curve ${\bf r}(s)$, $s$ measures the position along
the contour, $\beta=(k_BT)^{-1}$ and $k_BTl_p=\kappa$ specify the stiffness
of the chain. The inextensibility of the chain is expressed by the local
constraint $|{\bf\hat {t}}(s)|$=1. The distribution function
$\rho_s({\bf r},l_p)$ of end-to-end vector ${\bf r}$ of a segment of
contour length $l_b$ can be found from the average of $\xi(1,n)
\delta({\bf r}-{\int_0}^{l_b}{\bf\hat{t}}(s)ds)$ over all chain conformations.
Here $\xi(1,n)$ represents the boundary condition that ensures smooth
variation of tangent vectors at the two ends of the segment in dsDNA. In an
ensemble of segment of length $l_b$ in a dsDNA which in solvents
forms a fractal structure, we may relax the boundary condition
and approximate $\rho_s({\bf r},l_b)$ by the end-to-end
distribution function of a chain of length $l_b$ with a free end.

The distribution function $\rho_s({\bf r},l_b)$ has been found by
Monte Carlo simulation [21,22] and analytically using approximate
schemes [23,24,25]. For $l_b/l_p\leq1$,$\rho_s({\bf r},l_b)$
exhibits sharp peak at $r\sim l_b$ whereas for $l_b/l_p >1$ the peak
broadens and shifts to smaller values of $r/l_b$. For $l_b/l_p\geq1$
a simple analytical expression has been found \cite{23} which can be
written as
\ba
\rho_s(r,l_b)=N_c\left(1-\frac{r^2}{l_b^2}\right)^{-9/2}exp
\left(\frac{-3}{4}\frac{l_b}{l_p}\frac{1}{(1-r^2/l_b^2)}\right)
\ea
where $N_c$ is the normalization constant and its value
is found by the requirement
\ba
4\pi\int_0^\infty\rho(r,l_b)r^2dr=1
\ea
Eq.(9) reproduces quite accurately the simulation results for $l_b/l_p>1$
\cite{21}.

If  we use $\rho_s(r,l_b)$ given by Eq.(9)
for $\rho(r,l_b)$ and substitute into Eq.(2) we get for $l_b/l_p\geq 1$
\ba
Z_n(l_b)=\frac{K_2(l_b)}{l_b^3}
\ea
where
\ba
K_3(l_b)\simeq {\frac{1}{\gamma^{3/2}}
{\frac{4+12({\gamma\alpha})^{-1}+15({\gamma\alpha})^{-2}}
{4+12{\alpha}^{-1}+15{\alpha}^{-2}}\left(\frac{3}{4\pi a_0l_s}\right)^3}} ,
\ea
{\hspace{2cm}{\large$\alpha=\frac{3}{4}l_b/l_p $}} 
\hspace{2cm} and {\hspace{2cm}{\large$\gamma=1+\frac{2l_p}{a_0l_s}$}}

The value of $K(l_b)$ depends on the value of $l_b/l_p$. For
$l_b/l_p\gg 1$, $K_3(l_b)$ reduces to 
\ba
\left(\frac{3}{4\pi}\frac{1}{a_0l_s}\right)^3\left(\frac{1}{\gamma}\right)^{3/2}
\ea
 which agrees with the value given in (7). We note that while the exponent
$c$ remains $3$ the prefactor $K$ depends on the form chosen for
$\rho(r,l_b)$.

The inclusion of excluded volume interactions will decrease the number of
configurations found from the Gaussian chains forming the bubble. To see
this we consider the effect of self-avoiding interactions
within chains $C_1$ and $C_2$. The number of self-avoiding walks
(SAWs) which start at origin and arrive at a point
defined by the position vector ${\bf r}$ is given as \cite{16}
\ba
P_n({\bf r},l_0)\sim \mu^{l_0}l_0^{(\gamma-1-3\nu)}g(r/l_0^\nu)
\ea
where $\mu$ is the connectivity constant, $g(x)$ is a scaling
function, $\gamma$ is the entropic and $\nu$ is the metric exponent.
Though the function $g(x)$ is not exactly known for SAWs, it has been approximated
as $g(x)\sim x^{\phi}e^{-\lambda x^{\delta}}$ where $\lambda > 0$
,$\delta=\frac{1}{1-\nu}$ and $\phi$ can be expressed in terms of
known exponents [15,28]. For the random walks $\phi=0$ and $\nu=1/2$. As
the bubble is formed by two self-avoiding walks each of length
$l_0$ with common end points, the total weights of all
configurations of the bubble of which the two end points
are separated by distance $r$ are
\ba
Z_n({\bf r},l_0)\sim\mu^{2l_0}l_0^{2(r-1-3\nu)}H(r/l_0^\nu)
\ea
where

{\hspace{2cm.}}$H(r/l_0^\nu)\sim\left(\frac{r}{l_0}\right)^{2\phi}
exp\left[-2\lambda(r/l_0^\nu)^{\frac{1}{1-\nu}}\right]$

The integral in Eq.(2) is no longer Gaussian, but can be evaluated
using the steepest descent method. Here we, however, take simple view
and consider the limiting case of the bubble being embedded
in an infinitely long rod shaped dsDNA. As for this case $\rho
({\bf r},l_b)=\delta(r-l_b)/4\pi{l_b}^2$, the integral
is easily evaluated giving $c=2-2\nu+6\nu$. Using approximate
SAW exponents, $\gamma=1.158$, $\nu=0.588$ [16] we find
$c=3.2$. The analysis given here ignores the excluded volume
interactions between the two chains and between chains
 and the segments of dsDNA which inclusion will further
increase the value of $c$. A
more systematic treatment of the effect of excluded volume
interactions on $c$ will be given in a future publication.

In conclusion, we suggest that the partition function of the bubble 
embedded in a dsDNA should be calculated using expression given by Eq.(2).
There are two terms in this expression. First, the total weights
of possible configurations of a system of two ssDNA of given lengths which 
start from a point along the contour of dsDNA and reunite
at another point on dsDNA at a distance
$r$. The  points of origin and reuniting of two ssDNA
are the points where bubble connects with the intact dsDNA strands and are called
zipper forks. Second, the probability density of finding the zipper forks
of a bubble of given length
at distance $r$ apart. In the closed walks models which have been used 
to estimate the partition function of the bubble the second term has
been neglected. We calculate the total weights of possible configurations
of two chains forming the bubble by treating them to be 
noninteracting Gaussian chains. We also consider the case when 
they are represented by the self-avoiding walks model. For the
probability density of finding the zipper forks 
at distance $r$ apart in long dsDNA we suggest that it can be
approximated by the distribution function of end-to-end
vector ${\bf r}$ calculated for dsDNA using the wormlike chain
model. For the Gaussian chain model the exponent $c$
is found to be 3 and when self-avoidence in each chain were
considered its value increased to 3.2. Inclusion of the excluded volume
interaction between chains and between segments of dsDNA will further
increase the value of $c$.    

The work was supported by a research grant from DST of Govt. of India, New Delhi.


\begin{figure}[]
\includegraphics[width=4.5in]{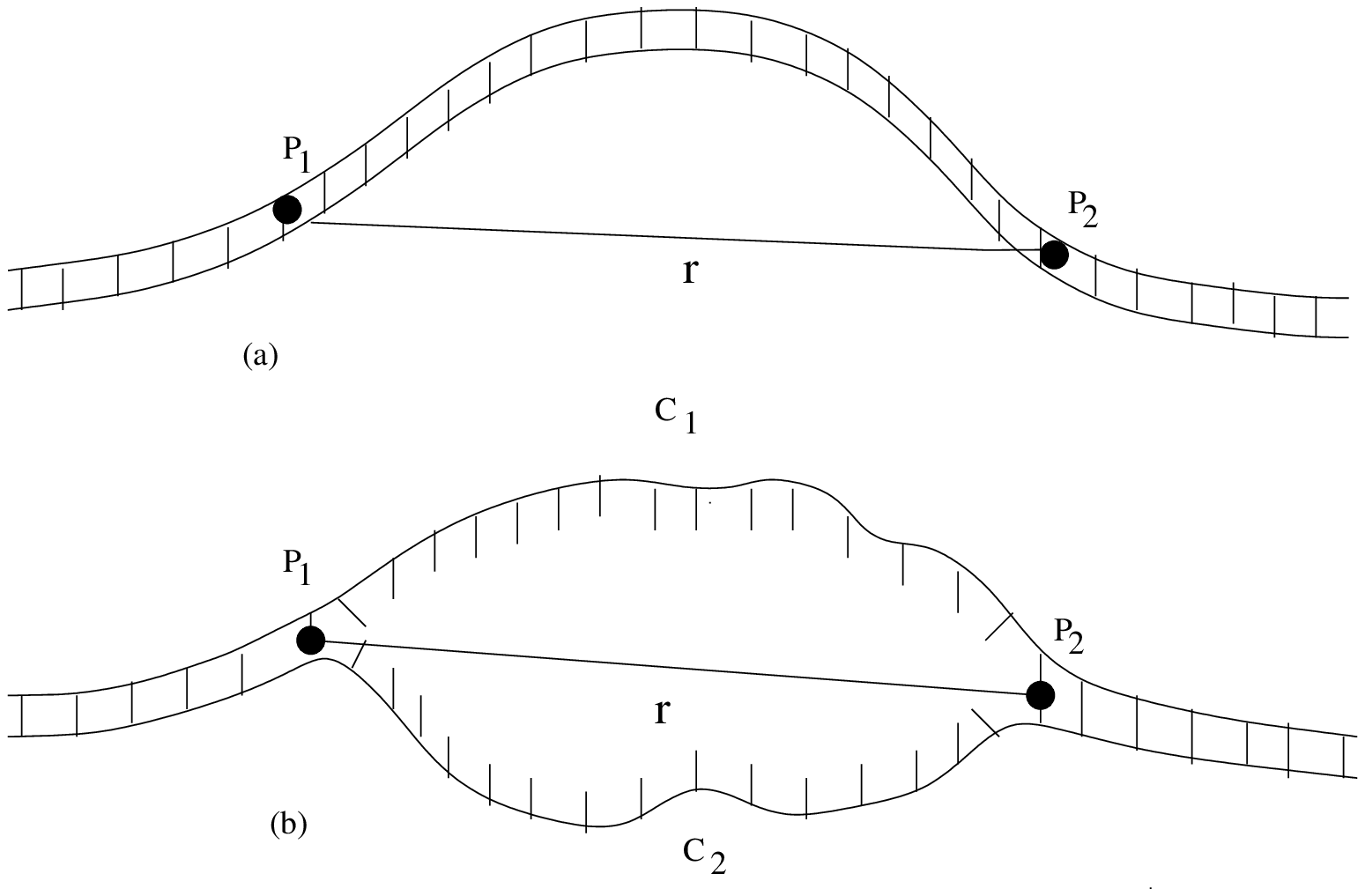}
\caption{Fig. 1(a) Shows a segment of dsDNA. The contour length
between points $P_1$ and $P_2$ which are separated by distance
${\bf r}$ is 3-d space is equal to $l_b=nb_0$ where $n$ is the
number of base-pairs and $b_0$=3.4$\AA$.
1(b) Shows bubble formed by unzipping of all $n$ base-pairs 
between points $P_1$ and $P_2$. The length of single 
stranded DNA $C_1$ and $C_2$ is equal to $l_0=na_0$ where
$a_0\simeq$ 6$\AA$. The points $P_1$ and $P_2$ represent the 
zipper-forks of the bubble. A zipper fork is junction point 
between dsDNA and the bubble.}
\end{figure}

\end{document}